\documentstyle[11pt,newpasp,twoside, epsfig, epsf]{article}
\begin{document}
\title{Backwards galaxy formation with only a few parameters}
\author{Ignacio Ferreras \& Joseph Silk}
\affil{Nuclear \& Astrophysics Lab. Keble Road, 
Oxford OX1 3RH, United Kingdom}

\begin{abstract}
We present a simple phenomenological model of
star formation in galaxies that describes the process with
a set of a few parameters. The star formation efficiency
($C_{\rm eff}$) and the fraction of gas and metals ejected
in outflows ($B_{\rm out}$) are assumed to be the main drivers 
of star formation. The Tully-Fisher and the Faber-Jackson relations
in different passbands are used as constraints in the analysis of 
disk and elliptical galaxies, respectively. We find that a steep 
correlation between $C_{\rm eff}$ and maximum rotation velocity is 
needed in disk systems, whereas elliptical galaxies can be explained by a 
high star formation efficiency, uncorrelated with the central velocity
dispersion.
Gas outflows are not important in disks, whereas
a correlation of outflows with galaxy mass must be invoked in early-type
galaxies
to account for the observed colors. A simple explanation for these
correlations can be given with respect to the difference in the
dynamical formation histories of disks and elliptical galaxies.
A phenomenolgical  model allows us to evolve the system ``backwards'' in
time.
A significant redshift evolution of the {\sl slope} of the 
Tully-Fisher relation is predicted, with the rest frame $B$-band slope
($\gamma_B=\Delta\log L_B/\Delta v_{\rm rot}$) steepening from 
$\gamma_B\sim 3$ at $z=0$ up to $4.5$ at $z\sim 1$.
\end{abstract}


\section{Introduction}
One of the main difficulties in the study of galaxy formation is
to determine the detailed role played by star formation. The complexity of 
this process both locally --- in the environment of the giant 
molecular clouds --- and globally --- via the importance 
of star formation in the structure and evolution of galaxies --- 
are major hurdles towards a final theory of galaxy formation. 
The remarkably tight
correlation between luminosity and maximum rotation velocity
found by Tully \& Fisher (1977) in disk galaxies has been 
the Holy Grail of galaxy formation modellers. Its origin has been
claimed to be in cosmology, and attributed to  the
structure of
dark matter halos (Mo, Mao \& White 1998; Steinmetz \& Navarro 1999).
However, star formation may play an important role (Heavens \& 
Jimenez 1999; Silk 2001) and the assumption of light tracing 
mass may not be correct (McGaugh et al. 2000).
In such a situation, a phenomenological approach seems appropriate. 
The process of star formation may be reduced to a simple set of
parameters, 
and high-quality observations are used as constraints. We 
concentrate on the scaling relations between luminosity and 
some ``dynamical'' parameter such as the maximum rotation velocity 
in disk galaxies (Tully \& Fisher 1977) or the  central velocity
dispersion 
in elliptical galaxies (Faber \& Jackson 1976). In this paper we 
compare the difference between the star formation histories
in early- and late-type galaxies. 


\section{Star Formation Phenomenology}
Our model of galaxy formation follows a ``backwards'' approach. 
We describe the process of star formation with a small number of
parameters.
For a given choice of parameters we integrate the chemical enrichment
equations to find a final distribution of stellar ages and metallicities
which are convolved using simple stellar populations from the latest
models of Bruzual \& Charlot (in preparation). The predicted 
spectrophotometric properties are compared to the observations, so
that we come up with a range of parameters that can explain the
observations. This phenomenological analysis is useful for the study 
of the physics of the underlying star formation.
Furthermore, it allows us to evolve the system backwards in time, so that
its predictions over a large redshift range can be used to further
constrain the volume of parameter space compatible with the observations.
The parameters used in this model can be separated into two categories:
external and internal
(see Ferreras \& Silk 2000; 2001 for a complete description):

\begin{figure}[t]
\begin{minipage}[b]{6.6cm}{
\psfig{figure=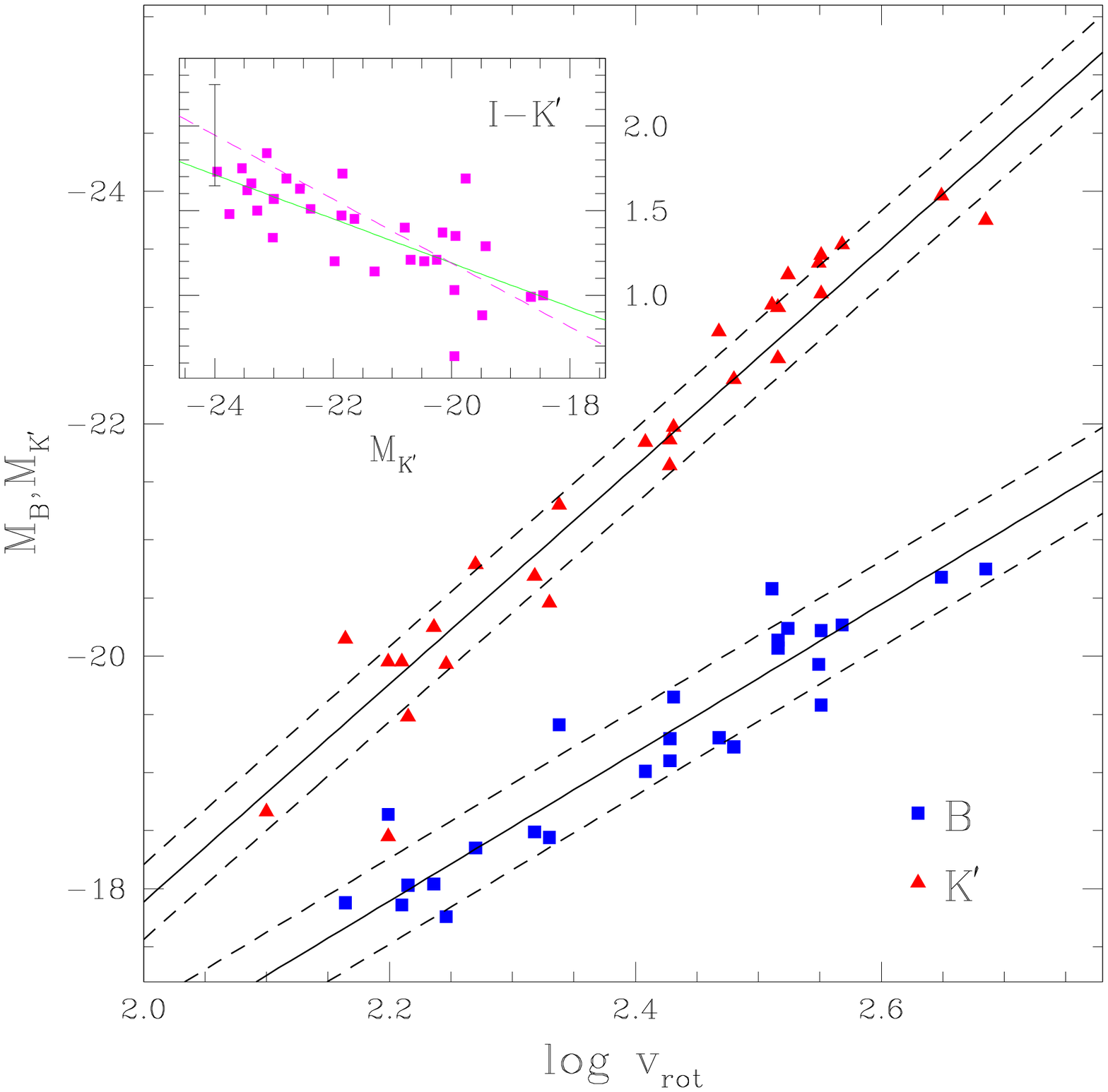,width=6.0cm} 
}
\end{minipage}
\begin{minipage}[b]{6.5cm}{
\psfig{figure=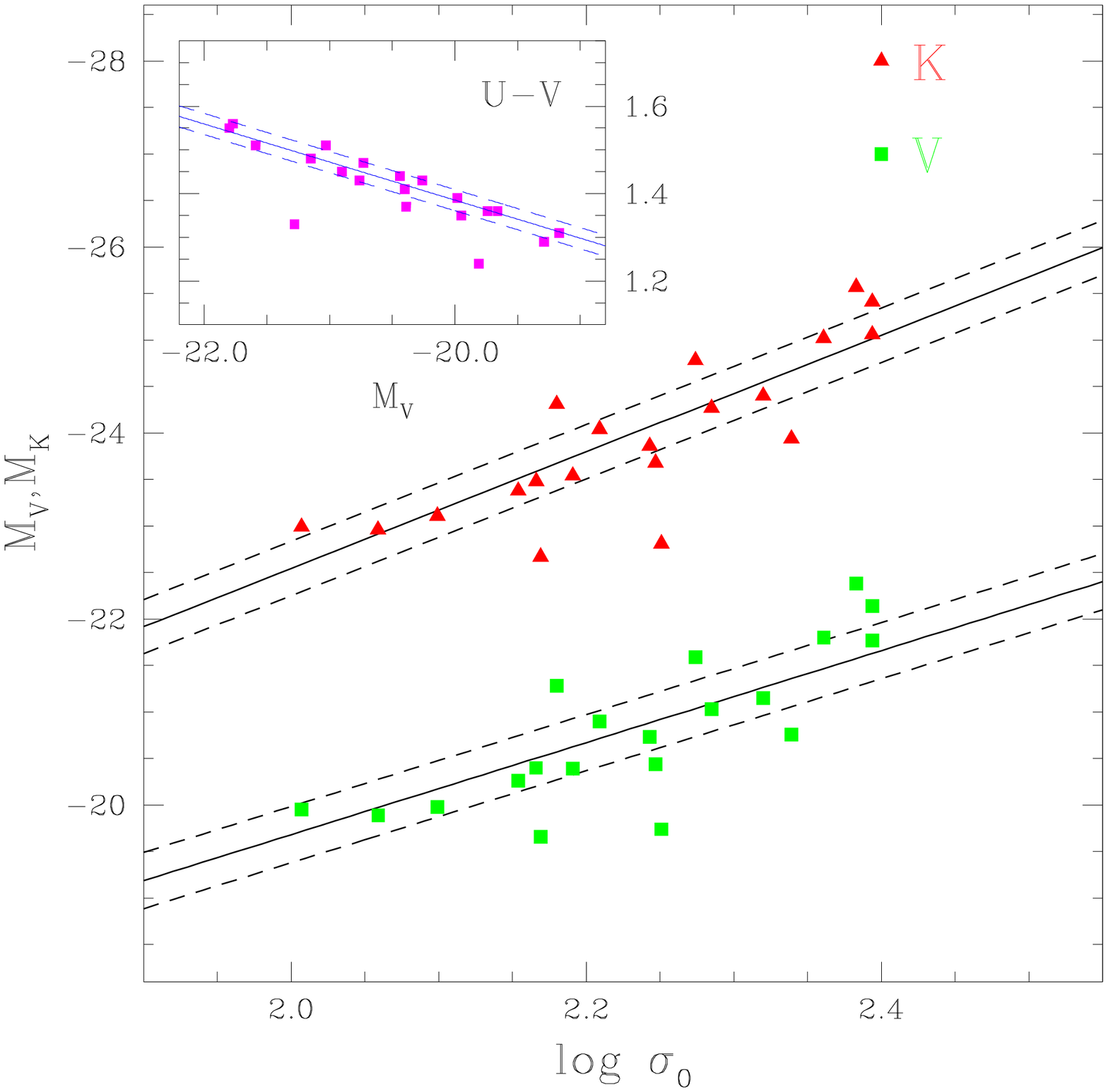,width=6.0cm}
}
\end{minipage}
\caption{Observational constraints: Multiband Tully-Fisher 
(left; Verheijen 1997) and Faber-Jackson relations (right; 
Bower et al. 1992) for disk and elliptical galaxies, respectively.
The solid lines are the linear fits used as constraint, along with
the scatter, represented by dahsed lines. The insets in both panels
are the expected color-magnitude relations as a consequence of the
different slope between bandpasses in either case. }
\end{figure}

\begin{itemize}
\item[$\bullet$] {\sl External}. These parameters control the infall
of primordial gas that fuels star formation. We define 
a short ``start-up'' timescale --- $\tau_1\sim 0.5$~Gyr --- which allows 
for the buildup of metals, preventing the over-production of old,
low-metallicity stars. The main external parameters are the infall
timescale --- $\tau_2$ --- which can be related to galaxy type, with
early-types having values $\tau_2\sim 1-2$~Gyr and late-types
having a more extended infall $\tau_2\sim 8-10$~Gyr, and the
epoch at infall maximum. We characterize 
this epoch by its redshift $z_F$. Even though this epoch could be
determined in a hierarchical clustering scenario, either using 
dynamical simulations or the analytic Press-Schechter formalism, 
we decided to leave this parameter unconstrained, to find out which 
formation epochs would be ruled out purely based on spectrophotometric
considerations.

\item[$\bullet$] The {\sl internal} parameters are responsible for 
the star formation rate and chemical enrichment. We define a star 
formation efficiency $C_{\rm eff}$ as the constant of proportionality
between the star formation rate ($\psi$) and a power of the gas mass 
density ($\rho_g$) as given by a Schmidt law: 
$\psi (t)= C_{\rm eff}\rho_g^n(t)$. This efficiency is roughly the
inverse of the time to process gas into stars (in Gyr in our units).
Semi-analytic models usually assume this efficiency to be given by the 
inverse of the dynamical timescale in the galaxy. The other internal
parameter is the amount of gas ejected in outflows ($B_{\rm out}$). 
It is still a matter of debate whether supernova-driven outflows are 
important in disk galaxies.
\end{itemize}

The observational constraints are shown in Figure~1. The 
multiband Tully-Fisher (Tully \& Fisher 1977) and Faber-Jackson 
(Faber \& Jackson 1976) relations are used to explore 
the star formation history of late-type and early-type galaxies, 
respectively. Figure~1 shows the data for some of the
observed bandpasses. The disk galaxy sample corresponds to a 
set of disks in the Ursa Major cluster observed by 
Verheijen (1997) in $B$, $R$, $I$ and $K^\prime$ bands.
The early-type galaxies used in this work are the sample of 
Coma ellipticals observed by Bower, Lucey \& Ellis (1992)
in $U$, $V$, $J$ and $K$ bands.

The slope difference between filters
can be represented in terms of a color-magnitude relation
{\sl both} in ellipticals and disk galaxies (see insets in 
Figure~1). However, the scatter found in disk galaxies
is much larger, notwithstanding the added complication 
of the contribution from dust in disks.
The universality of the color-magnitude
relation in disks is a reasonable assumption if we compare
the fit for a completely different sample of disk galaxies
observed by de~Grijs \& Peletier (1999) and shown in the
inset of Figure~1 ({\sl left}) as a dashed line. 
These authors also found similar slopes in several
sets of disk galaxies from the literature.
The small scatter found in ellipticals
even in high redshift clusters (Van~Dokkum et al. 2000) 
is an indication of the dominant role of metallicity 
to explain the color range in these galaxies. The bulk 
of the stellar populations in early-type systems must 
be rather old in order to explain this small scatter if
a fine-tuned conspiracy in the star formation history is to
be avoided. On the other hand, the large scatter found in
the color-magnitude relation of disk galaxies is a tell-tale
sign of a more extended period of star formation. 

The {\sl modus operandi} of the model presented here
is to select one star formation history by a choice of
parameters $(\tau_1,\tau_2,z_F,C_{\rm eff},B_{\rm out})$. This
history is integrated using the standard enrichment
equations (Ferreras \& Silk 2001) and the final chemical
enrichment tracks are used to convolve simple stellar populations
over a wide range of ages and metallicities from the latest
population synthesis models of Bruzual \& Charlot (in preparation),
out of which we come up with a spectral energy distribution
which gives us a set of broadband fluxes which are compared
with the data described above. This procedure is repeated
for different choices of parameters. Even though
this model is a coarse oversimplification of the star formation
process in galaxies, its strength is the robustness allowed by
exploring a wide volume of parameter space. In our model we do 
not assume {\sl a priori} any functional dependence of
the star formation efficiency, the outflow fractions or even
the epoch of maximum infall. The outcome of this model should
therefore be useful for  implementing  the ``recipes'' for galaxy 
formation included in semi-analytic models.

\begin{figure}[t]
\begin{minipage}[b]{6.6cm}{
\psfig{figure=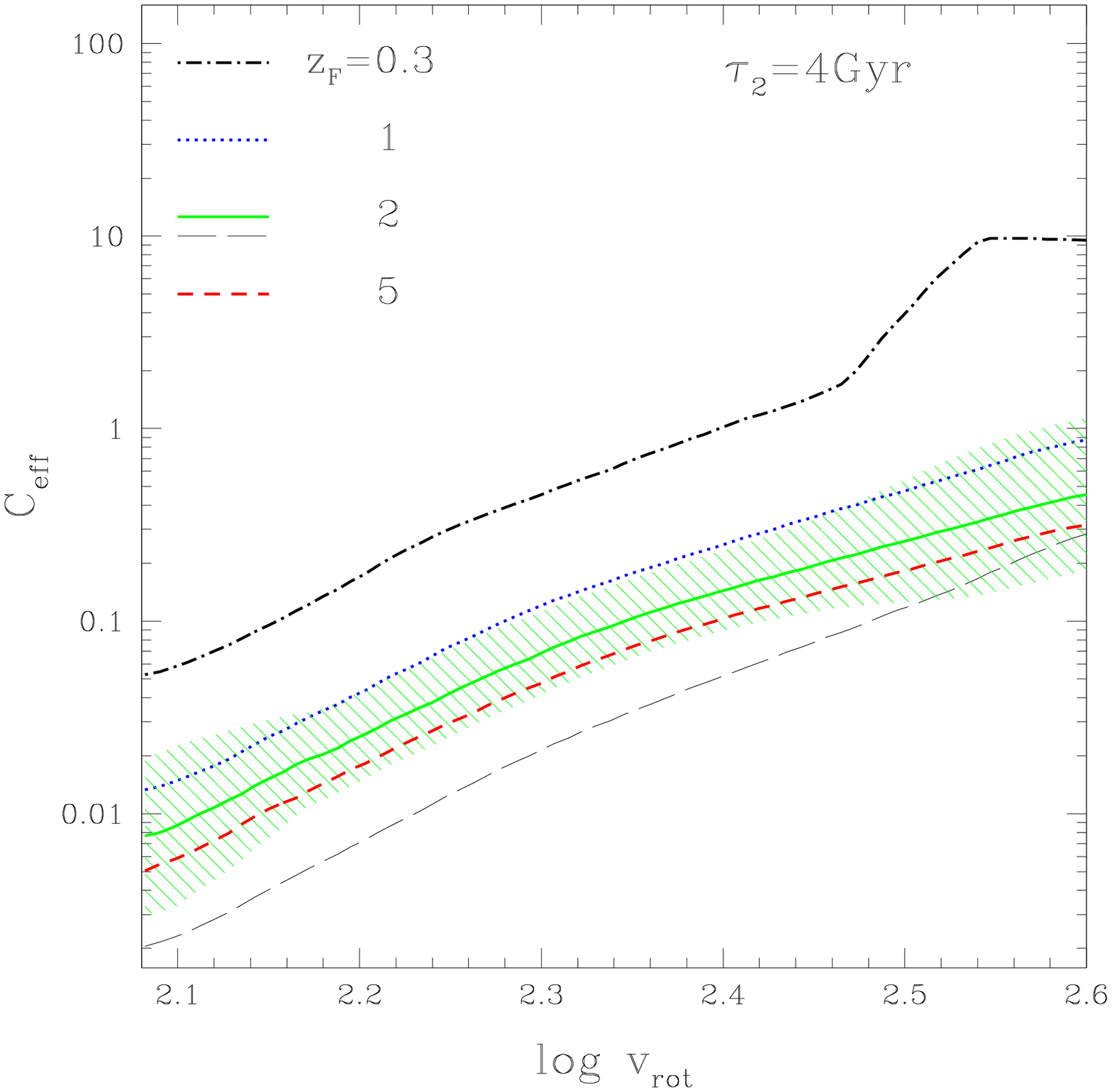,width=6.0cm} 
}
\end{minipage}
\begin{minipage}[b]{6.5cm}{
\psfig{figure=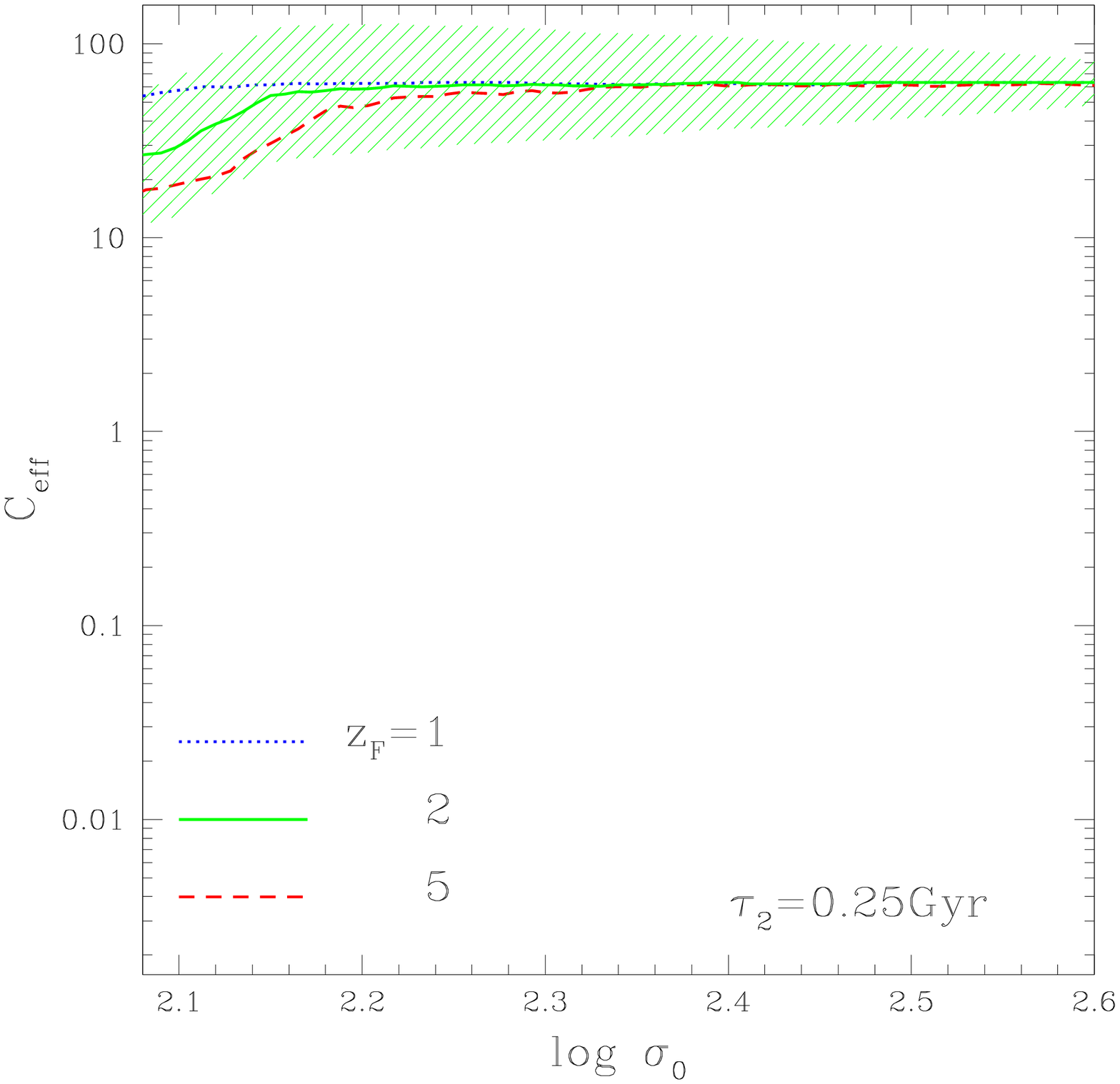,width=6.0cm}
}
\end{minipage}
\caption{Star formation efficiencies ($C_{\rm eff}$) compatible with 
the observations of disk ({\sl left}) and elliptical ({\sl right}) 
galaxies. The shaded area represents the 90\% confidence level in the 
$\chi^2$ test for the $z_F=2$ case. The thin long-dashed line in
the left panel corresponds to a quadratic Schmidt law for $z_F=2$. 
The other lines assume a linear law.}
\end{figure}


\section{The efficiency of Star Formation}
Figure~2 shows the range of star formation efficiencies predicted 
by the model. A $\chi^2$ test was applied to compare the predicted 
colors for a given choice of parameters and the linear fits to the
observed data shown in Figure~1. A typical infall timescale in 
early and late-type systems was chosen for these figures, but a 
different value of $\tau_2$ does not change the result significantly 
(Ferreras \& Silk 2001). A remarkable difference is found between 
the efficiency in disks and elliptical galaxies. The former have 
efficiencies that range over nearly two orders of magnitude, 
being strongly correlated with the maximum rotation velocity 
of the disk (or roughly the luminosity or even the disk mass). 
On the other hand, elliptical galaxies have very high star 
formation efficiencies which are not correlated with galaxy mass. 

This result should not come as a surprise if we realize the
colors of ellipticals are very red and the scatter of the 
color-magnitude relation is remarkably small. There is an
interesting point that can be predicted from the figure: The
high star formation efficiency required to explain the data
is telling us that pure passive evolution should be expected 
for early-type cluster galaxies with redshift. This has been
observed already in several samples of clusters at moderate
and high redshift (e.g. Stanford et al. 1998) but 
the constraint for the efficiencies shown in Figure~2 only 
involves zero redshift data from the Coma cluster. 

The large range of efficiencies found in late-type galaxies 
implies low mass disks should have a very high mass 
fraction of gas. The slope of the correlation between maximum
rotation velocity and star formation efficiency 
($C_{\rm eff}\propto v^\eta_{\rm rot}$) is $\eta\sim 3.5-4$
depending on infall parameters and whether a linear or a 
quadratic Schmidt law is used.
In order to check how inefficient star formation can be
(and so how high the gas mass fraction can be), we show in 
Figure~3 the colors predicted by our model, enforcing a
power-law dependence between $v_{\rm rot}$ and $C_{\rm eff}$
for a few choices of power law index. The result shows that
the correlation must be steep in order to reproduce the
colors. Correlations with $\eta\sim 1-2$ are incapable of
reaching blue colors for any choice of infall parameters.

\begin{figure}[t]
\begin{center}
\leavevmode
\psfig{figure=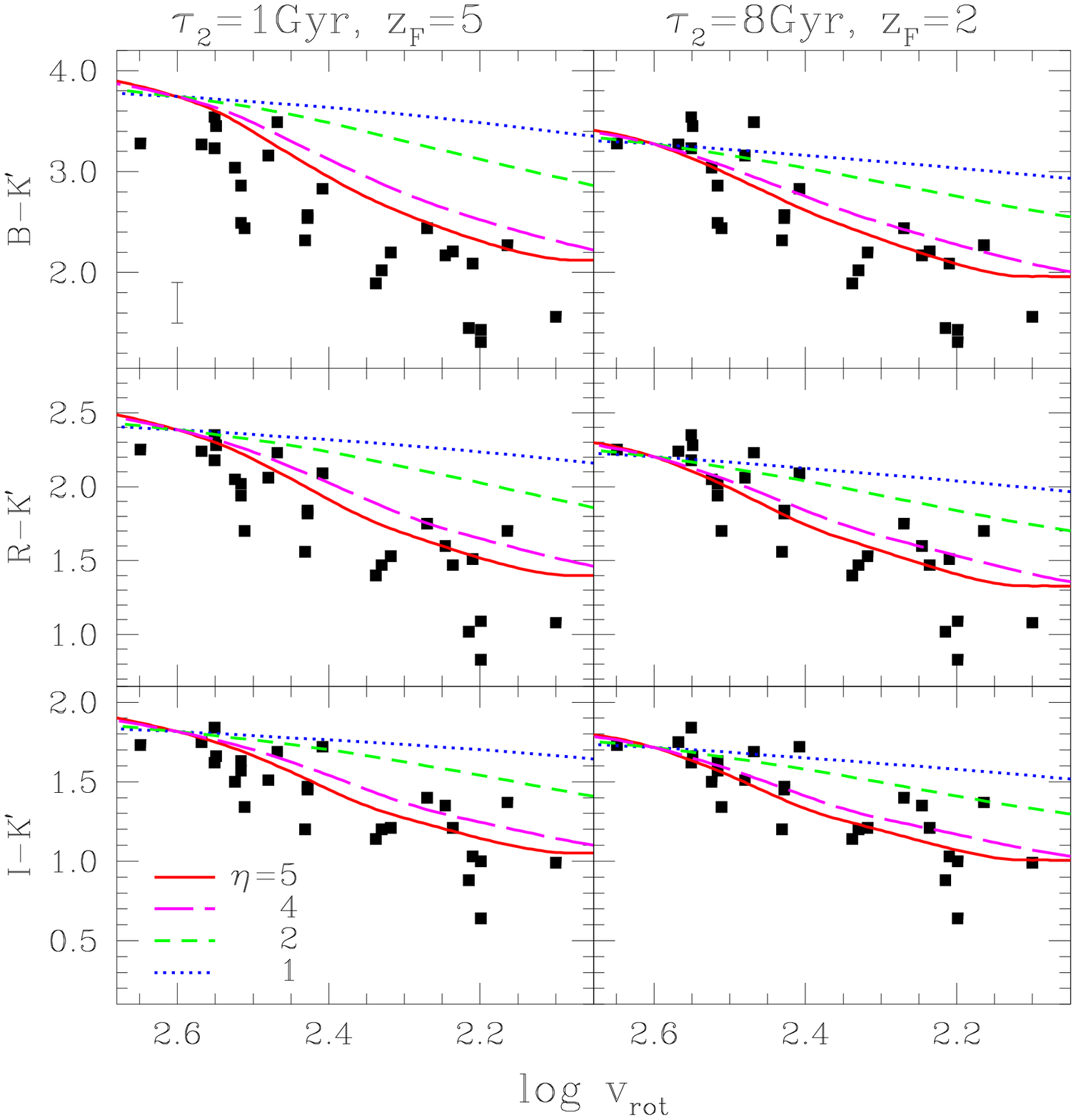,width=7cm} 
\caption{Predicted colors for a model which enforces the
star formation efficiency as a fixed power law with respect to
maximum rotation velocity: $C_{\rm eff}\propto v_{\rm rot}^\eta$.
The data points are from Verheijen (1997).}
\end{center}
\end{figure}

The steepness of the correlation was further analyzed in 
Ferreras \& Silk (2001) with respect to the predictions
of the star formation rates at zero redshift and compared 
with a local sample. The model gives consistent answers. 
This leads us to conclude that a large amount of molecular 
hydrogen is needed in order to explain the high gas mass 
fractions predicted for low mass disks. This is a highly 
speculative prediction of  our relatively simple model. However, 
the ``canonical'' method of estimating molecular hydrogen
using carbon monoxide as tracer (e.g. Young \& Knezek 1989)
may underestimate the contribution from $H_2$ in low metallicity
regions. Furthermore, {\sl direct} detections 
of large amounts of $H_2$ through its rotational transitions
(Valentijn \& Van der Werf 1999) as well as recent theoretical 
estimates of the lifetimes of the giant molecular clouds that nurture
star formation (Pringle et al. 2001) point to  the possible
existence of 
large masses of hitherto undetected, cold $H_2$. More work in this field 
is badly needed and the prospect of a partial solution to the baryonic 
dark matter problem via an augmentation in the cold gas mass
seems very
encouraging.
We note paranthetically that the warm/hot intergalactic medium predicted
from large-scale hydrodynamical  simulations
accounts for at most  half of the baryonic dark matter.

\begin{figure}[t]
\begin{minipage}[b]{6.6cm}{
\psfig{figure=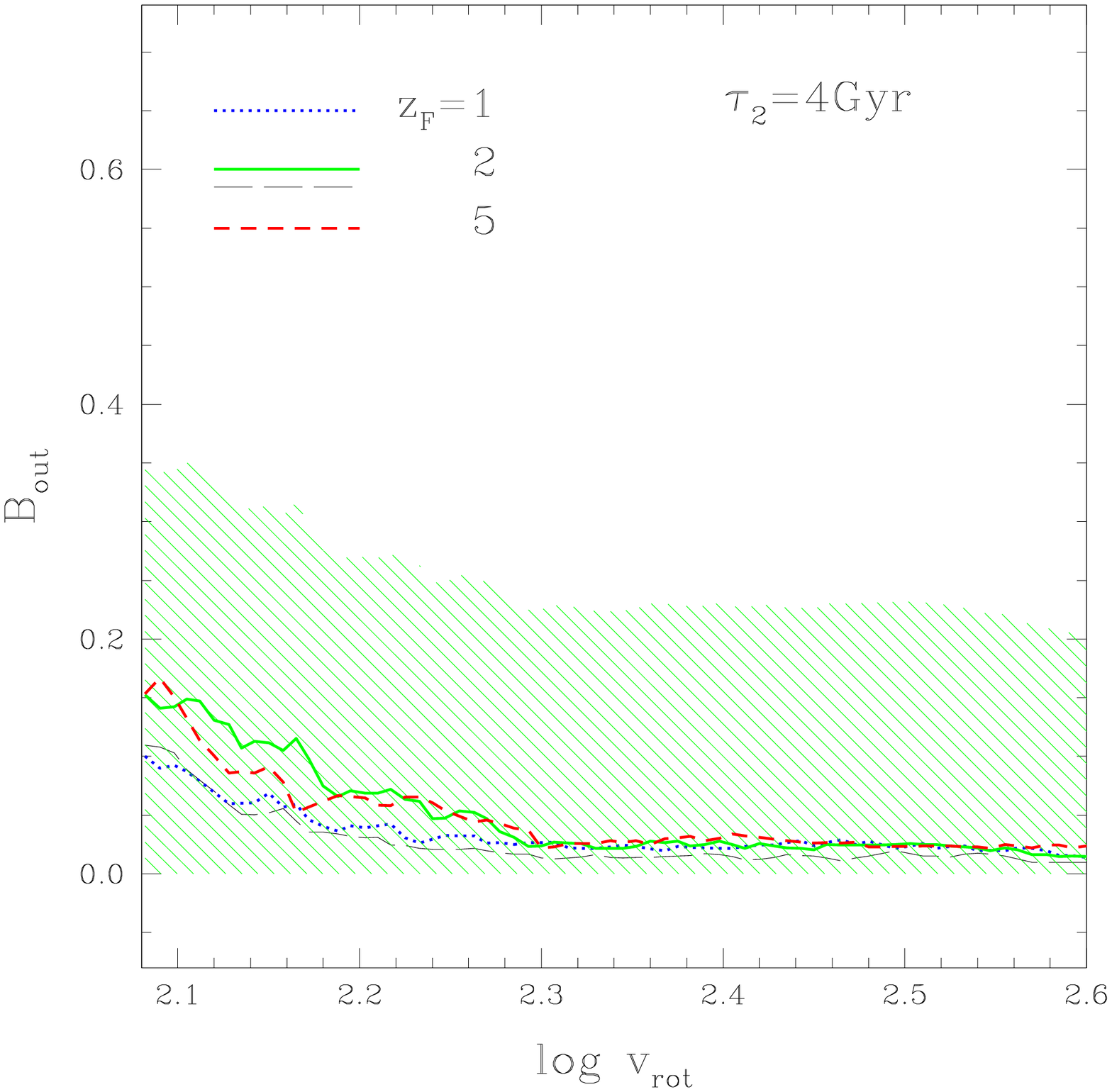,width=6.0cm} 
}
\end{minipage}
\begin{minipage}[b]{6.5cm}{
\psfig{figure=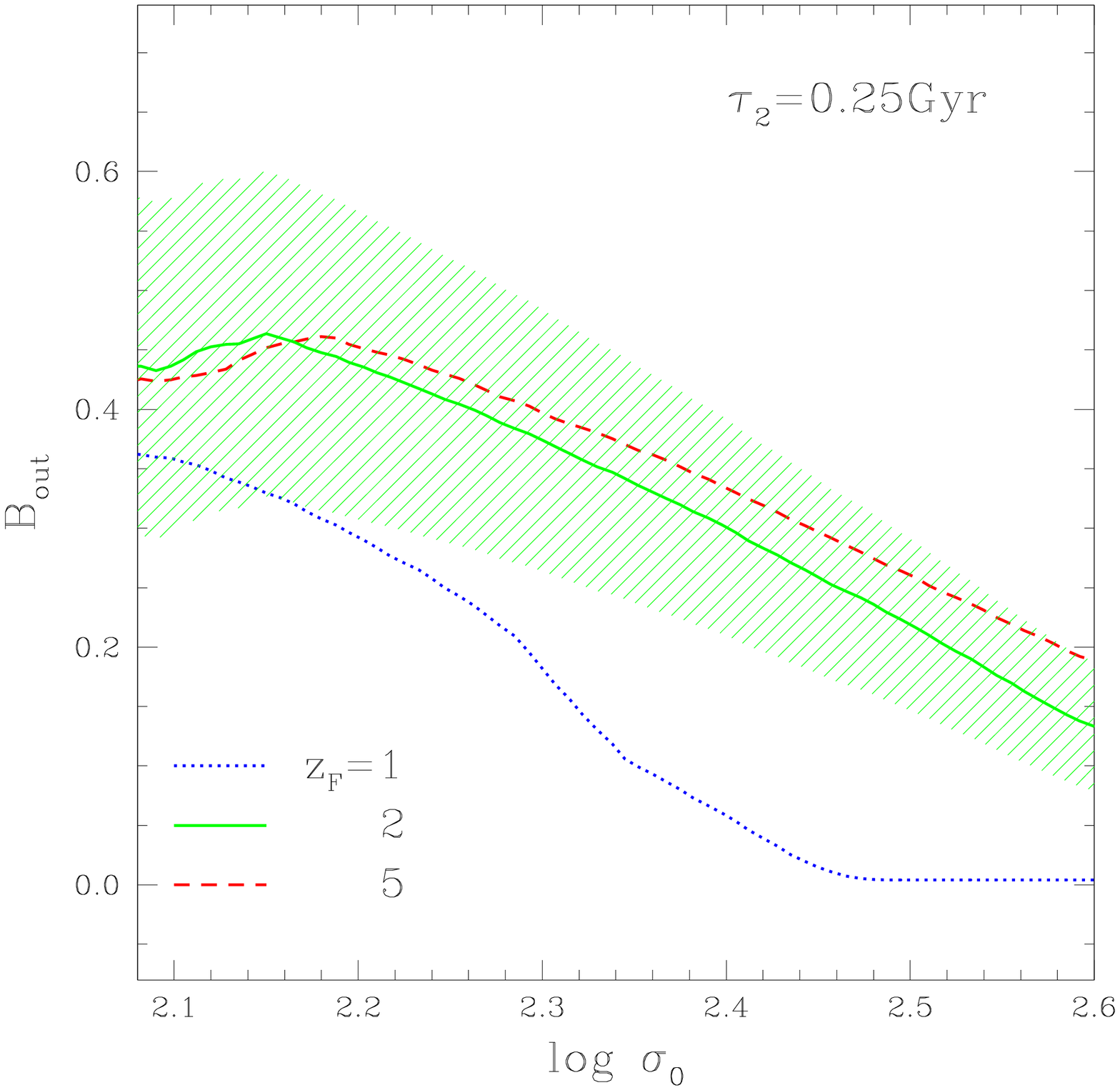,width=6.0cm}
}
\end{minipage}
\caption{Predicted range of outflows as a function of the dynamical
parameter in disk ({\sl left}) and elliptical ({\sl right}) galaxies. 
The shaded area gives the 90\% confidence level of the
$\chi^2$ test. The lines are coded in the same way as for the models
presented in figure~2.}
\end{figure}

\section{Gas Outflows}
The ejection of gas and metals in outflows changes the 
metallicity distribution of the stellar
populations. If the outflow fraction is very large, it could even
deplete the gas reservoir, ``stifling'' star formation. 
However, the average metallicities that would be found in
such a system would be very low so that the predicted photometry would
not be compatible with the observations. Nevertheless, our model 
explores a wide range of outflows $0\leq B_{\rm out}< 1$, using
only the observed colors to reject unphysical values.
Figure~4 shows the predicted outflow mass fractions 
both for disks ({\sl left}) and elliptical galaxies ({\sl right}).
One can see from the figure that gas outflows may not be important
in disk galaxies. The observations are compatible with no gas outflow
throughout the mass range. This result is in good agreement
with hydrodynamical simulations in disk galaxies 
(MacLow \& Ferrara 1999). The 2-dimensional geometry of disks
prevent a large fraction of gas from being ejected out of the
galaxy. Most of the gas ``stirred'' by supernovae-driven winds
stays in the interstellar medium. Only in small galaxies (well
below the range of maximum rotation velocities analyzed 
in this paper) feedback from supernovae seems to play an 
important role.

On the other hand, elliptical galaxies present an important 
range of outflows, being very significant at the
faint end of the luminosity function. This is necessary in order to obtain
the observed 
color range in cluster ellipticals. Figure~2 shows that early-type
systems have a high star formation efficiency. This translates into
rather old stellar populations. In this case, the only
way to explain a color-magnitude relation is to assume a mass-metallicity
relation. This is confirmed by the lack of evolution in the slope
of the color-magnitude relation in clusters up to redshifts $z\sim 1$
(Stanford et al. 1998). Hence, a mass-metallicity relation can only
be obtained by a correlation between mass and gas$+$metal outflows
as seen in the figure. We want to emphasize that the conclusions
reached here with respect to star formation efficiencies and
gas outflows are purely based on spectrophotometry and not on 
the dynamics of these galaxies. The fact that these results 
are compatible with the current ideas about the dynamical history
of early- and late-type systems should count as independent 
further evidence towards the galaxy formation scenario.


\section{The Dynamics/Star Formation connection}
Star formation is assumed to take place mostly (or even completely)
in giant molecular clouds. However, it is not clear which mechanisms 
are responsible for the onset of this process. Turbulence and 
magnetic fields play an important role in the process of
star formation. Nevertheless, a simple scenario for star 
formation involving cloud-cloud collisions seems to give reasonable
results (e.g. Tan 2000; Silk 2001). This implies the quiescent 
dynamics of disk
galaxies should result in very low star formation efficiencies. 
Furthermore, massive disk galaxies should have more cloud-cloud 
encounters compared to low mass disks, thereby raising the average 
star formation efficiency. On the other hand, the dynamically 
violent birth of elliptical galaxies after the
merging of progenitors with similar masses (Toomre \& Toomre 1972)
implies a higher rate of cloud-cloud collisions, yielding high 
star formation efficiencies that are uncorrelated with the 
progenitor masses. Hence, the dependence of the star formation
efficiency with respect to some ``mass-related parameter'' such as the
rotational velocity in disk galaxies or central velocity 
dispersion in elliptical systems can be explained --- at
least qualitatively --- with respect to the dynamical 
history of the galaxies. 

The dependence on galaxy mass of the outflow fraction ($B_{\rm out}$)
presented in Figure~4 can also be explained in terms of the
dynamical formation process of both galaxy types: disks are presumed
to form by an ``ordered'' infall of gas on to the center of dark
mater halos (e.g. White \& Rees 1978). This implies outflows can only be
triggered by supernova-driven winds, a process that seems not to
be too effective in the 2D geometry of disk galaxies, even with masses
at the faint end of the observed Tully-Fisher relation 
($v_{\rm rot}\geq 100$ km s$^{-1}$). Notice that this analysis
has not been extended to dwarf galaxies. In fact, dwarf galaxies
drop off the luminosity Tully-Fisher relation and are claimed
by McGaugh et al. (2000) to ``jump back'' to a linear correlation
if baryonic mass --- rather than luminosity --- is plotted 
against rotation velocity. Our analysis is purely based on 
luminosity and so we would have to add the non-linearity of the
low-mass end of the Tully-Fisher relation to draw any conclusions
regarding dwarf galaxies.
On the other side of the morphological spectrum, early-type 
systems are thought to have been formed in merging processes 
whose progenitors had similar masses. 
This mechanism involves a violent encounter with low
relative velocities, and it is very efficient in driving
gas, metals and even stars out of the galaxy. Hence, the
results shown in Figure~4 are qualitatively consistent with 
the dynamical history of early- and late-type galaxies.

A quantitative estimate is harder to get. Wyse \& Silk
(1989) proposed a star formation efficiency that scales with 
the local angular frequency: $C_{\rm eff}(r)\equiv\epsilon\Omega (r)$.
Tan (2000) uses a simple model of cloud-cloud collisions to 
find a similar scaling behavior for the efficiency.
In our notation, their claim implies a global (i.e. spatially
integrated) efficiency of $C_{\rm eff}\propto v_{\rm rot}^2$.
The correlation shown in Figure~2 is significantly steeper 
than this, and only a steep dependence of disk sizes would
reconcile both estimates (however, see \S6).

Concerning gas outflows, high resolution Hydro$+$N-body
numerical simulations of mergers for a wide range of encounters
should be performed in order to estimate the scaling of the
outflow fraction $B_{\rm out}$ as a function of galaxy mass.


\section{Caveat: Formation Epoch}
One of the caveats of the model can be seen in Figure~2. The
star formation efficiency is shown as a dot-dashed line for an
extremely low formation redshift ($z_F=0.3$). In that case, a
similar correlation is predicted for $C_{\rm eff}$ vs
$\log v_{\rm rot}$ but the values are one order of magnitude
higher~!. This implies that a disk formation scenario
in which small disks form {\sl much later} than large disks
could result in a star formation efficiency that is completely
uncorrelated with the galaxy mass. This result has been suggested
by several authors (Bell \& Bower 2000; Boissier et al. 2001). 
According to a standard hierarchical
formation scenario, small disks should be dynamically 
older than their massive counterparts. However, one can think 
of several processes by which either infall or star formation
can be delayed in small disks. Even though we do not
believe values of $z_F\sim 0.1$ are realistic for our low
mass disks ($v_{\rm rot}\sim 100$ km s$^{-1}$), we must
emphasize that an inverted hierarchy in the formation of stars
in disks may make the predicted correlation between disk
mass and efficiency shallower. 

\begin{figure}[t]
\begin{center}
\leavevmode
\psfig{figure=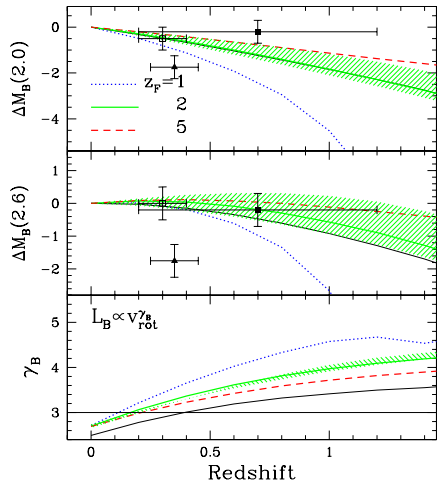,width=7cm} 
\caption{Redshift evolution of the $B$-band Tully-Fisher relation. The
top and middle panels show the magnitude evolution at $\log v_{rot}=2$
and $2.6$, respectively. The bottom panel gives the slope evolution.
The thick dotted, solid and dashed lines correspond to $z_F=1, 2$ and 
$5$, respectively (linear law). The thin line corresponds to 
$z_F=2$ for a quadratic Schmidt-law. See \S7 for details.}
\end{center}
\end{figure}

\section{Redshift evolution}
The model presented here allows us not only to connect 
phenomenology with physical processes driving
star formation. It also enables us to evolve the ``best fit''
for a given mass backwards in time, so that we can trace the
redshift evolution of color and luminosity (of course aside 
from the problem of dust formation and depletion). One of the
main consequences of the star formation efficiencies found in 
Figure~2 is that:
\begin{itemize}
\item[a)] (Unsurprising) The high and uncorrelated star formation 
	efficiency found in early-type galaxies implies their stellar 
	populations are old and so the luminosity evolution corresponds
	to a passively evolving simple stellar population. 
	The color-magnitude relation should not change its slope
	with redshift. Only the zero-point offset should evolve
	in step with the luminosity evolution of a simple stellar
	population (e.g. Bower et al. 1992; Stanford et al. 1998).
\item[b)] (Somehow surprising) The strongly correlated efficiency
	found in disk galaxies implies a huge difference between
	the stellar age histogram of low mass disks and their
	massive counterparts. This translates into a {\bf slope
	change} of the Tully-Fisher relation with redshift.
\end{itemize}

Figure~5 shows the redshift evolution of the Tully-Fisher relation
observed in rest frame $B$ band. The bottom panel plots the
evolution of the slope $\gamma_B$ (so that 
$L_B\propto v_{\rm rot}^{\gamma_B}$). Needless to say, a change in 
the slope will result in a different zero-point offset
between $z=0$ and $z>0$ depending on the absolute luminosity
of the galaxies observed. The top and middle panels show this
offset for low ($\log v_{\rm rot}=2$) and high mass ($2.6$) 
disks, respectively. The points are observed offsets 
of the Tully-Fisher relation at moderate redshifts: Bershady et al.
(1999, hollow square); Vogt (2000, filled square) and Simard
\& Pritchet (1998, triangle). The horizontal error bar encompasses the
redshift range of the observations. The horizontal line --- 
at $\gamma_B=3$ --- in the bottom panel gives the prediction of 
a simple dynamical model that assumes the same mass-to-light ratio
for all disks (Mo et al. 1998).


\section{Final Remarks}
A phenomenological model using only observed data as a constraint
enables us to determine the role of the star formation efficiency 
and gas outflows. The passband-dependent slope of both the Tully-Fisher
and the Faber-Jackson relations for disk and elliptical
galaxies, respectively yields a remarkably different picture
of star formation. Taking the model at face value, one could
venture into the risky business of extrapolation and consider that
the model described so far has only taken into account star formation.
Were we to determine the scaling between $C_{\rm eff}$ and 
$v_{\rm rot}$ from a so far undeveloped theory of star formation, 
we could have inferred that self-regulated star formation could
explain the Tully-Fisher relation (e.g. Silk 2001).

\acknowledgements
IF is supported by a grant from the European Community under 
contract HPMF-CT-1999-00109. We would very much like to thank
the organizers for such an enjoyable workshop.

\end{document}